\documentclass[12pt]{article}

\usepackage{amsmath,amsfonts,amssymb}

\newcounter{tabl}
\newcommand{\be}{\begin{equation}}
\newcommand{\ee}{\end{equation}}
\newcommand{\beq}{\begin{eqnarray}}
\newcommand{\eeq}{\end{eqnarray}}
\newcommand{\bea}[2]{\be\label{#2}\begin{array}{#1}}
\newcommand{\eea}{\end{array}\ee}

\def\Rb{{\rm \bf R}}
\def\Cb{{\rm \bf C}}

\def\Tr{\,{\rm Tr}\, }
\def\det{\,{\rm det}\, }

\def\({\left(}
\def\){\right)}
\def\[{\left[}
\def\]{\right]}
\def\p{\partial}

\def\11{1\!\! 1}

\def\hf{{1\over 2}}

\def\eps{\varepsilon}

   \def\CG {{\cal G}}
   \def\CH {{\cal H}}

   \def\CM {{\cal M}}

   \def\CP {{\cal P}}

   \def\CS {{\cal S}}

   \def\CV {{\cal V}}

\newcommand{\im}{\beta}

\newcommand{\tE}{\lefteqn{\smash{\mathop{\vphantom{<}}\limits^{\;\sim}}}E}
\newcommand{\tP}{\lefteqn{\smash{\mathop{\vphantom{<}}\limits^{\;\sim}}}P}
\newcommand{\tQ}{\lefteqn{\smash{\mathop{\vphantom{<}}\limits^{\;\sim}}}Q}

\newcommand{\Pt}{\lefteqn{\smash{\mathop{\vphantom{\Bigl(}}\limits_{\sim}
\atop \ }}P}
\newcommand{\Qt}{\lefteqn{\smash{\mathop{\vphantom{\Bigl(}}\limits_{\sim}
\atop \ }}Q}

\newcommand{\SA}{{\cal A}}
\newcommand{\SSA}{{\bf A}}

\newcommand{\gl}{{\rm g}}

\newcommand{\SLC}{{\rm SL(2,\Cb)}}
\newcommand{\sgchi}{${\rm SU}_{\chi}(2)$ }

\newcommand{\Hkin}{\CH_{\rm kin}}

\newcommand{\Ref}[1]{(\ref{#1})}

\renewcommand{\theenumi}{\roman{enumi}}
\renewcommand{\labelenumi}{(\theenumi)}

\def\LL{L}
\def\unity{{\lefteqn{1}\,1}}

\begin{document}
%
%

\title{
Spin foam model
from canonical quantization}

\author{Sergei Alexandrov\thanks{email: Sergey.Alexandrov@lpta.univ-montp2.fr}
}

\date{}

\maketitle

\vspace{-1cm}

\begin{center}
\emph{Laboratoire de Physique Th\'eorique \& Astroparticules} \\
\emph{Universit\'e Montpellier II, 34095 Montpellier Cedex 05, France}


\end{center}

\vspace{0.1cm}

\begin{abstract}
We suggest a modification of the Barrett-Crane spin foam model of
4-dimensional Lorentzian general relativity motivated by the canonical quantization.
The starting point is Lorentz covariant loop quantum gravity.
Its kinematical Hilbert space is found as a space of the so-called
projected spin networks.
These spin networks are identified with the boundary states of a spin foam model
and provide a generalization of the unique Barrette-Crane intertwiner.
We propose a way to modify the Barrett-Crane quantization procedure
to arrive at this generalization:
the $B$ field (bi-vectors) should be promoted not to generators of
the gauge algebra, but to their certain projection. The modification is also justified
by the canonical analysis of Plebanski formulation.
Finally, we compare our construction with other proposals to modify
the Barret-Crane model.

\end{abstract}

\thispagestyle{empty}

\newpage

\pagenumbering{arabic}

\tableofcontents

\section{Introduction}

The spin foam approach to quantum gravity provides a possibility to describe the dynamics
of quantum spacetime in a covariant and background independent way \cite{Oriti,perez}.
It allows to represent transition amplitudes as Feynman sums over two-complexes (branched surfaces)
colored with representations and intertwiners of a symmetry group.
Such complexes are called spin foams and the symmetry group is usually taken to be the gauge group
of the corresponding classical theory.
Every single spin foam can be viewed as a particular history of a quantum space represented
by a boundary state. The boundary states can be identified as spin networks appearing as graphs
with a similar coloring. Such kind of objects also arises as quantum states in the loop approach
\cite{loops1,loops2,Rov-dif,Rovbook} and it was shown that, applying the evolution operator to
the spin networks, one gets qualitatively the same picture as in spin foam models \cite{resrov}.
Moreover, in the Euclidean 3d case the precise agreement was found \cite{Noui:2004iy} where a matrix
element of the Hamiltonian of 3d gravity was shown to reproduce the vertex amplitude
of the Ponzano--Regge model \cite{PonzanoRegge}.

However, in four dimensions a quantitative agreement between the loop and the spin foam approaches
has not been established so far. The present situation can be summarized as follows.
On the spin foam side, the most popular and the most studied model for the 4-dimensional general relativity
is the model suggested by Barrett and Crane (BC), which exists in both Euclidean \cite{BCE} and Lorentzian
\cite{BC} versions. It can be obtained from the spin foam quantization of, respectively,
SO(4) or \SLC\ BF theory by appropriately restricting the representations and intertwiners labeling the
branched surfaces.

On the canonical loop side one has two options.
The first one is provided by the standard loop quantum gravity (LQG) \cite{Rov-dif,Rovbook}
based on the Ashtekar--Barbero canonical formulation
\cite{Ashtekar:1986yd,Ashtekar:1987gu,Barbero:1994an,Barbero:1994ap}.
This formulation has SU(2) as its structure group
and therefore there are little chances to recover the BC model starting from LQG.
Besides, the results of LQG are known to depend on the so called Immirzi parameter \cite{imir},
whereas there are no sign of this dependence in the BC model \cite{Mon,Liv}.

The second option, which is at our disposal, is much better in this respect.
It is given by the so called covariant loop quantum gravity (CLQG) \cite{Livine:2006ix}, which
is a loop quantization of a Lorentz covariant formulation developed in \cite{SA}.
As the BC model, it uses the full Lorentz gauge group and produces results independent of
the Immirzi parameter. Moreover, there is a close relation between the states of CLQG and
the BC boundary states. The latter turn out to be a particular subset of the former \cite{AlLiv,AK}.
Thus, at the kinematical level the two approaches {\it almost} agree, but still not precisely.

At the same time, there are various evidences that the BC model is not the correct spin foam
quantization of general relativity. Most of them can be traced back to the way this model
is obtained from BF theory. This is achieved by imposing the so called simplicity constraints
which turn the topological BF theory into non-topological 4-dimensional gravity.
Several indications point towards the conclusion that the specific procedure used by Barrett and Crane
imposes them too strongly and, as a result, the state space is too restrictive
\cite{Livine:2007vk,Engle:2007uq}. This gives rise to a hope that modifying this procedure
one can arrive to a model fully consistent with the canonical approach \cite{ABR}.

Recently, two such modifications have been proposed \cite{Livine:2007vk,Engle:2007uq}.
Both of them suggest to impose the constraints only on averages, weakly.
As a result, the models have a different set of boundary states and give rise to new vertex
amplitudes. In particular, the states in the model of \cite{Engle:2007uq} resemble
SU(2) spin networks of LQG, thus establishing a possible link with this canonical approach.
However, the two new models treat only the Euclidean case (in fact, the paper \cite{Livine:2007vk}
deals only with BF theory and results for general relativity have not been presented)
and the generalization to the Lorentzian case is not so trivial.
Besides, we will show that the quantization procedure used in \cite{Engle:2007uq} is not unique
and, actually, another quantization seems to be more preferable. Thus, the problem
of correct implementation of the simplicity constraints in the spin foam approach remains
still open.

Here we approach this problem using as a guideline the results obtained in the framework of CLQG,
namely the structure of states in its kinematical Hilbert space. Therefore, first,
we reconsider the construction of this space (see earlier works \cite{SAhil,AlLiv,Livine:2006ix}).
It requires to address a problem similar to the one existing in the spin foam approach:
how to impose second class constraints at the quantum level? In the first part of the paper
we give a solution to this problem obtaining a basis in the Hilbert space in terms of the so
called projected spin networks \cite{psn}. Assuming that there exists a spin foam model
consistent with our construction, one can find intertwiners associated with edges of
two-complexes (tetrahedra of a dual 4-simplex).
They generalize the unique Barrett--Crane intertwiner, thus confirming the necessity
to modify the BC model and providing an indication what the generalization should be.

Then we identify a weak point in the BC quantization procedure which is,
in our opinion, the assignment of
quantum operators to classical bi-vectors associated to triangles of a 4-simplex.
Usually, they are represented as generators of the gauge algebra. But we argue that
the proper way of treating the second class constraints appeals for another assignment:
a projection to a certain subspace determined by the normal to a tetrahedron must be inserted.
In this way all simplicity constraints turn out to be automatically solved.

The new assignment allows to relax the BC conditions on intertwiners, but it is not sufficient to
reproduce the intertwiners following from the canonical quantization. To this end,
we also relax the closure constraint allowing for an intertwiner to depend on the normal
to the tetrahedron. The resulting boundary states coincide with the projected spin networks
of the canonical approach.

Then we show that the quantization procedure proposed for a 4-simplex can be
generalized in a natural way, which has a direct correspondence on the canonical side.
In particular, there exists a quantization leading to a spin foam model
which is in precise agreement with LQG. Besides, it represents a Lorentzian
version of \cite{Engle:2007uq} with a few important differences.
But we argue that there are several reasons
why one expects problems with this quantization,
whereas the quantization consistent with CLQG is more preferable.

Finally, we discuss how the constructed boundary states can then be
used as an input for the vertex amplitude to be
associated to the quantum 4-simplex. We arrive at the conclusion that they are not enough
as the usual prescriptions used in spin foam models are not applicable. A more involved analysis
of the path integral is required which is beyond the scope of this paper.

\vspace{0.2cm}

{\bf Note added:} After submission of our paper, a very nice and interesting work \cite{Freidel:2007py}
has appeared. It proposes a new spin foam model, based on the ideas of \cite{Livine:2007vk},
for both Euclidean and Lorentzian general relativity
and simultaneously argues that the model of \cite{Engle:2007uq} describes the topological sector
of Plebanski formulation of gravity. As we will mention, some results of \cite{Freidel:2007py}
agree with our proposals, but we leave a more detailed comparison for a future work.

\section{Covariant loop quantum gravity}

The CLQG is a program of quantizing gravity {\it \`a la} loops
starting form a Lorentz covariant canonical formulation \cite{Livine:2006ix}.
This formulation was developed in \cite{SA} and further elaborated in \cite{SAcon,AlLiv,AK}.
In this section we review the main details of the classical framework (sec. \ref{subsec_phase})
and present a construction of the kinematical Hilbert space.
This is done in two steps. First, we study an enlarged Hilbert space
obtained by a quantization of the classical phase space with a symplectic structure
given by Dirac brackets (sec. \ref{subsec_enlarge}).
And at the second step, we impose second class constraints at the level
of the Hilbert space (sec. \ref{sec_kinH}).

\subsection{The phase space}
\label{subsec_phase}

The Lorentz covariant canonical formulation of \cite{SA} appears as the
Hamiltonian formulation of the following generalized
Hilbert--Palatini action \cite{Holst:1995pc}
\be
S_{(\beta)}=\frac12 \int
\eps_{\alpha\beta\gamma\delta} e^\alpha \wedge e^\beta \wedge
(\Omega^{\gamma\delta}+
\frac{1}{\beta}\star\Omega^{\gamma\delta}),
\label{palat}
\ee
where $e^\alpha$ is the tetrad field, $\Omega^{\alpha\beta}$ is
the curvature of the spin-connection $\omega^{\alpha\beta}$ and
the star operator is the Hodge operator defined as $ \star
\Omega^{\alpha\beta}=\frac 12 {\eps^{\alpha\beta}}_{\gamma\delta}
\Omega^{\gamma\delta} $. The parameter $\beta$ is called Immirzi
parameter and is not physical since the additional term in the
action vanishes on-shell and does not change the equations of
motion.

The resulting structure of the canonical formulation can be summarized as follows.
The phase space of the theory is parameterized by the fields
$\tP^i_X$, $\SA^X_i$ and $\omega^a$. To explain their relation to the original
tetrad and spin-connection, one needs to introduce some definitions.

First, $\tP^i_X$ is built of the components of the tetrad appearing
under the $3+1$ decomposition
\be
e^0=Ndt+\chi_a E_i^a dx^i ,\qquad e^a=E^a_idx^i+E^a_iN^idt.
\label{tetrad}
\ee
Here $E_i^a$ is the triad and $\chi_a$ describes deviations of the normal to the 3 dimensional
hypersurfaces foliating spacetime from the $t$-direction.
The particular case $\chi=0$ corresponds to the usual `time gauge' often used in the literature.
The requirement that the foliation is spacelike
is equivalent to the condition $\chi^2<1$ which will be assumed in the following.

Introducing the inverse triad and its densitized
version $\tE^i_a =h^{1/2}E^i_a$ with
$\sqrt{h}=\det E^a_i$, the field $\tP^i_X$ is defined as
\be
\tP_X^{ i}=(\tE^i_a,{\eps_a}^{bc}\tE^i_b\chi_c).
\label{multP}
\ee
The index $X$ can be thought as an antisymmetric pair of Lorentz indices $[\alpha\beta]$.
Then the first 3 components correspond to $[0a]$ and the other 3 are obtained
after contraction of $[ab]$ components with $\hf \eps^{abc}$.
It is convenient also to define a Hodge dual field
\be
\tQ_X^{ i}=(-{\eps_a}^{bc}\tE^i_b\chi_c,\tE^i_a)
\label{multQ}
\ee
and their inverse, $\Pt_i^X$ and
$\Qt_i^X$, which satisfy the following relations
\be
\tP^i_X=\Pi_X^Y\tQ^i_Y,
\qquad
\Pi_X^Y=\left(
\begin{array}{cc}
0&-1 \\ 1&0
\end{array}
\right)\delta^{ab},
\label{relPQ}
\ee
\be
\tQ^i_X\Qt_j^X=\delta^i_j,  \qquad
\tP^i_X\Pt_j^X=\delta^i_j,
\qquad
\tQ^i_X\Pt_j^X=\tP^i_X\Qt_j^X=0.
\ee
We refer to \cite{SA} for their explicit expressions.

A special role is played by the projectors
\be
I_{(P)X}^Y = \tP^{ i}_X\Pt_i^{Y},
\qquad
I_{(Q)X}^Y = \tQ^{ i}_X\Qt_i^{Y}
\label{proj}
\ee
satisfying
\be
\label{projpr}
I_{(P)X}^Y +I_{(Q)X}^Y=\delta_X^Y,
 \qquad
I_{(P)X}^Y \tP^i_Y=\tP^i_X, \qquad
I_{(P)X}^Y \tQ^i_Y=0,
\ee
and similar relations for $I_{(Q)}$ and the inverse fields.
These projectors depend only on $\chi$ and have a geometric
interpretation. Let us first consider the case $\chi=0$
corresponding to the `time gauge'.
Then $I_{(Q)}$ and $I_{(P)}$ project to the canonically embedded
$su(2)$ subalgebra of $sl(2,\Cb)$ and boosts from its orthogonal completion, respectively.
On the other hand, any non-vanishing $\chi$ satisfying $\chi^2<1$
can be obtained from the previous case by applying a boost.
Therefore, in a general case, $I_{(Q)}$
projects to a `boosted' subalgebra $su_{\chi}(2)$ obtained from the canonically
embedded one by applying the boost corresponding to $\chi$. The projector $I_{(P)}$
selects the orthogonal completion of the `boosted' subalgebra
(see \cite{AlLiv} for details).

These projectors will be crucial for the following construction.
As we will see, in the spin foam approach they will manifest as spacelike and timelike projectors
allowing to characterize the bi-vectors assigned to the triangles of a 4-simplex.

The other two variables, $\SA^X_i$ and $\omega^a$, parameterizing the phase space
originate from the spin-connection.
To start with, let us rewrite its space components as
\be
A_i^{ X}=(\omega_i^{0a},\frac12 {\eps^a}_{bc}\omega_i^{bc}).
\label{multA}
\ee
However, this connection is not suitable for quantization \cite{AV}
and instead a modified connection is considered
\be
\SA_i^X = \left(1+\frac{1}{\im^2}\right) I_{(P)Y}^X (R^{-1})^Y_Z A_i^Z
+ R^X_Y \Gamma_i^Y,
 \label{spconnew}
\ee
where
\be
\Gamma_i^X = \frac12 f^{W}_{YZ}I_{(Q)}^{XY} \Qt_i^Z \p_l \tQ^l_W
+\frac12 f^{ZW}_Y\left((\Qt\Qt)_{ij} I_{(Q)}^{XY}
+\Qt_j^X\Qt_i^Y -\Qt_i^X\Qt_j^Y \right) \tQ^l_Z \p_l \tQ^j_W
\label{gam}
\ee
is the \SLC\ connection on the 3d hypersurface compatible with the induced metric,
$f_{XY}^Y$ are $sl(2,\Cb)$ structure constants, $R^{X}_{Y} =\delta^{X}_{Y}-\frac{1}{\beta}\Pi^{X}_{Y}$
and their properties can be found, for example, in appendix A of \cite{AK}.

In contrast to the initial connection $A_i^X$,
the quantity \Ref{spconnew} has only 9 independent components due to
the relation
\be
I_{(Q)Y}^X\SA_i^Y=\Gamma_i^X(\tQ).
\label{contA}
\ee
Where are the other 9 missing components? First, one must take into account that
the variables $\tP^i_X$ and $A_i^X$ are subject to the second class constraints:
\beq
\phi^{ij}&=&\Pi^{XY}\tQ^i_X\tQ^j_Y=0,
\label{phi} \\
\psi^{ij}&=&2f^{XYZ}\tQ_X^{l}\tQ_Y^{\{ j}\partial_l \tQ_Z^{i\} }
-2(\tQ\tQ)^{ ij }\tQ_Z^{l}A_l^Z+
2(\tQ\tQ)^{l\{i  }\tQ_Z^{j\}}A_l^Z = 0.
\label{psi}
\eeq
Thus, 6 of the missing components disappear due to \Ref{psi}.
The other three must be taken as independent variables
and we denote them by $\omega^a$.

Besides the second class constraints,
there are ten first class constraints, $\CG_X,\ H_i$ and $H$, acting on this phase space.
They generate local Lorentz transformations, space and time diffeomorphisms, respectively.
They encode all dynamics of the theory because the Hamiltonian weakly vanishes being
a linear combination of the constraints.
We do not present their explicit expressions since they will not be needed in this paper.
The only important fact which is relevant for us here is that the quantity
\Ref{spconnew} transforms as a true Lorentz connection under all gauge symmetries.

The symplectic structure of the phase space
can be described by the Dirac brackets of our variables.
The commutation relations with $\tP^i_X$ and $\SA_i^X$ take the following form
\beq
\{ \tP_X^i(x),\tP_Y^j(y)\}_D&=&0, \nonumber \\
\{ \SA_i^X(x),\tP_Y^j(y)\}_D&=&\delta_i^j I_{(P)Y}^X \delta(x,y),
\label{comm} \\
\{ \SA^X_i(x), \SA^Y_j(y)\}_D& =&
\CM_{ij}^{XY} \delta(x,y), \nonumber
\eeq
where $\CM_{ij}^{XY}$ is a complicated $\im$-independent linear differential operator.
We refer to \cite{AK} for its explicit expression.
An important consequence of \Ref{comm} is
\be
\{ \SA^X_i,I_{(P)}^{YZ}\}_D=0.
\label{chicomm}
\ee
It demonstrates that the field $\chi_a$ commutes with both $\tP$ and $\SA$.
As a consequence, its conjugated field must be $\omega^a$,
the only one which does not commute with $\chi_a$.
On the other hand, $\chi_a$ transforms under the action of the Gauss constraint
$\CG_X$. (Recall that it can be actually used to parameterize the boosts.)
Hence, $\CG_X$ cannot be expressed only in terms of $\tP$ and $\SA$, but also
contains $\omega^a$ (cf. the canonical analysis of the Hilbert--Palatini action
in \cite{Alexandrov:1998cu}).
This fact, allows to avoid to work explicitly with $\omega^a$.
In particular, we do not need to write its Dirac brackets because it will appear
only in the combination $\CG_X$ whose commutators with both $\tP$ and $\SA$
are explicitly known.\footnote{One can think that the variable $\omega^a$ is removed
by solving the boost part of the Gauss constraint. But it is important to keep
track of $\omega^a$ when one discusses transformations under gauge symmetries.}

To conclude, we have a phase space parameterized by $\tP_X^i$, $\SA_i^X$
(and $\omega^a$ encoded in $\CG_X$)
with the symplectic structure given by \Ref{comm}.
The variables are subject to the second class constraints.
In particular, the connection $\SA_i^X$ satisfies \Ref{contA},
which can be thought as a representation of these constraints.

The representation \eqref{contA} is very convenient because it allows immediately to conclude
that the rotational part of the connection is non-dynamical being completely
determined by the triad and the field $\chi$. This is reflected, for example,
in the commutator with $\tP^i_X$
\eqref{comm} since its rotational part vanishes.
This situation is to be contrasted with the one which leads to the (Lorentz generalization of the)
Ashtekar--Barbero formulation. In that case the connection satisfies \cite{AlLiv}
\be
I_{(P)Y}^X \SSA_i^Y=\Pi^X_Y \Lambda_i^Y(\chi), \qquad
\Lambda_i^X=\(-\frac{\eps^{abc}\chi_b\p_i\chi_c}{1-\chi^2},\frac{\p_i\chi^a}{1-\chi^2}\).
\label{su2con}
\ee
Thus, for a constant $\chi_a$ its boost part is fixed to zero.
The fact that the only non-trivial components of the connection
are from the SU(2) subgroup enormously facilitates the quantization (which however suffers from anomalies
due to the failure of $\SSA_i^X$ to produce the correct time diffeomorphism transformations \cite{SAcon,AlLiv}).
In our case the constraint \eqref{contA} will also play an important role
in the construction of the Hilbert space.

\subsection{Enlarged Hilbert space and projected spin networks}
\label{subsec_enlarge}

The quantization of a dynamical system defined on the phase space
described in the previous subsection is complicated by the presence of the second class
constraints \Ref{contA}. Although they are already taken into account by means of
the Dirac brackets, they lead to a huge non-physical degeneracy in the state space.
Therefore, it is necessary to impose them also at the level of the Hilbert space.
This is a complicated problem and it will be addressed in the next subsection.
Here we review and add some new details to the construction of the so called
enlarged Hilbert space, which represents a quantization of our phase space
ignoring the second class constraints \cite{AK}. Then the physical
Hilbert space must be obtained by imposing the constraints on this enlarged space.

Following the idea of the loop quantization \cite{loops1,loops2},
we consider functionals of the connection $\SA$ associated with graphs
embedded into a 3-dimensional space. However, due to the property
\Ref{chicomm}, one can add $\chi$ to the list of configuration variables.
Thus, our state space is represented by functions of both $\SA$ and $\chi$.
This feature makes our situation different from the standard case appearing, for example, in LQG.

To introduce a structure of the Hilbert space on these states,
one considers generalized cylindrical functions \cite{psn}.
As the usual ones, they are defined on a graph and depend on the connection
through holonomies along its edges. At the same time,
they depend also on the values of the field $\chi$ at each node of the graph.
Since $\chi$ gives rise to an element
\be
x(\chi)=\(\frac{1}{\sqrt{1-\chi^2}},\frac{\chi^a}{\sqrt{1-\chi^2}}\)
\label{xxx}
\ee
of a hyperboloid in $\CM_4$, which is the homogeneous space
$X\equiv {\rm SL}(2,\Cb)/{\rm SU}(2)$, a generalized cylindrical function
associated with a graph $\Gamma$ with $E$ edges and $V$ vertices
is defined by a function on $[\SLC]^E\otimes [X]^V$.
It can be represented as
\be
F_{\Gamma,f}[\SA,\chi]=
f\(U_{\gamma_1}[\SA],\dots,U_{\gamma_E}[\SA];x(\chi(v_1)),\dots,x(\chi(v_V))\),
\label{pcf}
\ee
where $U_{\gamma_k}[\SA]=\CP \exp \(\int_{\gamma_k} \SA \)$
is the holonomy along the $k$th edge giving an element
of the Lorentz group.
The invariance under local Lorentz transformations imposes
the following invariance property
\be
f( \gl_{t(1)} g_1 \gl_{s(1)}^{-1},\dots,\gl_{t(E)}g_E \gl_{s(E)}^{-1};
 \gl_1\cdot x_1,\dots, \gl_V\cdot x_V)=f(g_1,\dots,g_E;x_1,\dots,x_V),
\label{invprop}
\ee
where $t(k)$ and $s(k)$ denote, respectively,
the target and the source vertex of the $k$th edge of the graph $\Gamma$,
$\gl_v \in \SLC$ and its action on $x_v$ coincides with the usual Lorentz transformation.
The gauge invariant scalar product of two generalized cylindrical functions
is given by \cite{psn}
\be
\langle F_{\Gamma,f}|F_{\Gamma',f'}\rangle =
\int_{[\SLC]^E} \prod\limits_{k=1}^E
d g_k \, \overline{ f (g_1,\dots,g_E;x_1,\dots,x_V )}
f' (g_1,\dots,g_E;x_1,\dots,x_V ),
\label{scpr}
\ee
where the variables $x_v$ are fixed to arbitrary values.
The invariance of the Haar measure ensures that the scalar product
does not depend on their choice.
Since the set of all generalized cylindrical functions is dense
in the space of all smooth gauge invariant functionals of $\SA$
and $\chi$, the enlarged Hilbert space $\CH_0$ can be obtained as
the completion of the space of these functions
in the norm induced by the bilinear form \Ref{scpr}.

A basis in $\CH_0$ can be found by performing the harmonic analysis
of the functions defined on $[\SLC]^E\otimes [X]^V$ and subject to the
condition \Ref{invprop}.
It turns out that there are two ways to resolve this condition.
To facilitate the solution, let us first impose the invariance
under the boosts. They can be used to put all $x_v$ to the origin
corresponding to the time gauge $\chi=0$. As a result, one
remains with functions on $[\SLC]^E$ which satisfy
\be
f( h_{t(1)} g_1 h_{s(1)}^{-1},\dots,h_{t(E)}g_E h_{s(E)}^{-1})
=f(g_1,\dots,g_E),
\label{invpropSU}
\ee
where $h_v\in{\rm SU}(2)$. It is clear that a basis in the space of
these functions is realized by \SLC\ spin networks such that
every edge $\gamma_k$ is endowed with a unitary irreducible representation
$\lambda_k=(n_k,\rho_k)$ of \SLC\
from the principal series\footnote{These are representations
appearing in the decomposition of a square integrable function on \SLC\
and labeled by a non-negative half-integer $n$ and by a real number $\rho$.}
and at every vertex $v$ one has an
intertwiner ensuring the invariance \Ref{invpropSU}. Now the question
is: what is the space of these intertwiners?

\subsubsection{The first basis}

Let $k(v)$ labels the edges meeting at the vertex $v$. Then the space of
intertwiners associated to this vertex coincides with SU(2)
invariant subspace in the representation space given by the direct product
of $\CH_{\SLC}^{\lambda_{k(v)}}$. The easiest way to pick up such an invariant
subspace is to decompose the product into a direct integral of irreducible
representations, then to decompose each of them into a direct sum
of irreducible representations of the SU(2) subgroup and finally to select only
the trivial representations:
\beq
\label{repsp}
& \displaystyle{\mathop{\bigotimes}\limits_{k(v)} \CH_{\SLC}^{\lambda_{k(v)}}}
=\displaystyle{\int d\lambda_{v}
\, N\(\lambda_{v},\{\lambda_{k(v)}\}\)\,\CH_{\SLC}^{\lambda_{v}}} = &
\\
&  \displaystyle{\mathop{\bigoplus}\limits_{n_{v}} \int d\rho_{v}
\, N\((n_{v},\rho_{v}),\{\lambda_{k(v)}\}\)
\(\mathop{\bigoplus}\limits_{j_v\ge n_{v}} \CH_{\rm SU(2)}^{j_v}\)
\longrightarrow
\int d\rho_v\, N\((0,\rho_v),\{\lambda_{k(v)}\}\) \,\CH_{\rm SU(2)}^0}, &
\nonumber
\eeq
where we denoted by $N\(\lambda_v,\{\lambda_{k(v)}\}\)$ the degeneracy of the representation
$\lambda_v=(n_v,\rho_v)$ appearing in the initial decomposition. Since the trivial
representation is present in the decomposition of only the so called simple
representations for which $n_v=0$, only those contribute
to the direct integral of the invariant subspaces. Thus, we arrive at the conclusion
that the \SLC\ intertwiners ensuring the SU(2) invariance are in one-to-one correspondence
with the Lorentz invariant intertwiners between the representations $\lambda_{k(v)}$
and a simple representation $\lambda_v=(0,\rho_v)$.

As a result, the basis elements of the enlarged Hilbert space are labeled
by graphs with the following coloring:
\begin{enumerate}
\item  \SLC\ irreducible representations assigned to the edges;
\item \SLC\ simple irreducible representations assigned to the vertices;
\item \SLC\ invariant intertwiners assigned to the vertices, which couple
representations of both types.
\end{enumerate}

The dependence of the basis elements on $\chi$ can be trivially restored by undoing
the boost transformations. Notice that the appearance of the simple
representations at vertices is quite natural because they are only ones which
are present in the decomposition of functions on the homogeneous space $X$. Thus,
if one first performs the harmonic analysis for the initial functions on
$[\SLC]^E\otimes [X]^V$ and imposes the invariance condition \Ref{invprop}
afterwards, one obtains the same result.

\subsubsection{The second basis}

There is however another way to proceed. Instead of doing like in \Ref{repsp},
one can first decompose each representation $\lambda_{k(v)}$ into a direct sum
of irreducible representations of the subgroup and only then to decompose
the product and to select the invariant subspace:
\be
\mathop{\bigotimes}\limits_{k(v)} \CH_{\SLC}^{\lambda_{k(v)}}
=\mathop{\bigotimes}\limits_{k(v)}
\(\mathop{\bigoplus}\limits_{j_{k(v)}\ge n_{k(v)}} \CH_{\rm SU(2)}^{j_{k(v)}}\)
=\mathop{\bigoplus}\limits_{j_{v}} N\(j_{v},\{ j_{k(v)}\}\)
\,\CH_{\rm SU(2)}^{j_{v}}
\longrightarrow  \mathop{\bigoplus}\limits_{n=1}^{N\(0,\{ j_{k(v)}\}\)}
\CH_{\rm SU(2)}^0.
\label{repsp2}
\ee
Following this procedure, one obtains that the basis of SU(2) invariant \SLC\
intertwiners can be labeled by  SU(2) irreducible representations,
associated to every incoming (or outgoing) edge and subject to the restriction
$j_k\ge n_k$, and by invariant SU(2) intertwiners between these
representations. Since an edge connects two vertices,
in total one associates two representations, $j_{t(k)}$ and $j_{s(k)}$,
to every edge:
the first representation is attached to the final point
of the edge, and the second one corresponds to its beginning.

To summarize and to compare with the previous case,
the elements of the second basis are labeled by graphs carrying:
\begin{enumerate}
\item  \SLC\ irreducible representations assigned to the edges;
\item two SU(2) irreducible representations assigned to the edges, which
appear in the decomposition of the corresponding \SLC\ representation;
\item SU(2) invariant intertwiners assigned to the vertices.
\end{enumerate}

From \Ref{repsp2} it is also clear how to construct the basis
functions. One should take holonomies along the edges in the representations
$\lambda_k$, to act on them by projectors to
$\CH_{\rm SU(2)}^{j_{t(k)}}$ and $\CH_{\rm SU(2)}^{j_{s(k)}}$
from the left and right, respectively, and finally to couple
the resulting objects using the intertwiners.
Remarkably, this procedure can also be done
in a Lorentz covariant way what allows to restore the dependence on $\chi$.
The only modification necessary to include a non-vanishing $\chi$
is that the projection must be done on the representations of a boosted
subgroup \sgchi, which was introduced in the discussion of the projectors
after eq. \Ref{projpr}.

As it is easy to realize, the constructed basis elements are precisely
the projected spin networks introduced in \cite{psn}. Initially they were
found as an example of gauge invariant generalized cylindrical functions,
which are eigenstates of the area operator for the surfaces intersecting
the spin network only at vertices.\footnote{They represent a refined version of
the spin networks `projected at each point' of \cite{SAhil} where
the gauge covariant projection to the subgroup was first introduced.}
Here we also proved that they realize
an orthogonal basis in the enlarged Hilbert space.

Notice that the projected spin networks form only one of the
two natural bases. At first sight, the two bases look quite different. In particular,
we trade a continuous label in the first one for a set of discrete labels in
the second one. It would be interesting to understand their relation in more detail.
In the following we will be concerning only with the second basis
since the projected spin networks are particularly useful in establishing relations
with other approaches.

\subsection{Kinematical Hilbert space of CLQG}
\label{sec_kinH}

\subsubsection{Preliminaries}

Now we are going to implement the second class constraints
on the enlarged Hilbert space constructed in the previous subsection.
But first, let us explain on a simple example why one needs to care about
constraints which have been already taken into account by means of the Dirac bracket.

Let we quantize a system with the phase space parameterized by canonical coordinates
$(q_a,x)$ and momenta $(p_a,y)$. The coordinate $x$ is supposed to be compact.
The only non-vanishing commutation relations
are $\{q_a,p_b\}=\delta_{ab}$ and $\{x,y\}=1$. Besides, we assume that the system is
subject to the second class constraints $x=y=0$. It is clear that the corresponding
Dirac bracket ensures that $x$ and $y$ commute with any function and does not change other
commutation relations. Quantizing the theory, we choose to work in the
coordinate representation, but for the variable $x$ we pass to the analog of the loop basis.
Thus, the elements of the Hilbert space are functions $\psi_j(q,x)=f(q)e^{ijx}$ which carry an index $j$
labeling representations of U(1). However, all physical operators involve only
$q_a$ and $p_a$ and do not act on the index $j$. Also the momentum $y$ vanishes on all
states so that one of the second class constraints is ensured.
As a result, the label $j$ does not carry any physical
information and, to describe the Hilbert space of the quantized system, it is
enough to restrict ourselves to a subset with a fixed value of $j$. Moreover,
regarding the second class constraint $x=0$, the natural value for $j$ is zero
since it is the only value for which $\psi_j(q,x)=\psi_j(q,0)$.
Note, however, that expectation values of physical observables
do not depend on the chosen value of $j$.

This example shows that the second class constraints should be imposed
to remove a degeneracy in the state space coming from degeneracy of the Dirac bracket.
In an appropriate loop basis, they may be taken into account by restricting
some labels of spin networks in such a way that, first, expectation values of
physical observables do not depend on the fixed labels and, secondly,
restriction to the constraint surface does not change the wave functions.

\subsubsection{Imposing second class constraints}
\label{sec_impsec}

Thus, our problem is to understand how the condition \Ref{contA}
can be implemented as a restriction on
the coloring of the projected spin networks, which will reduce
$\CH_0$ to the kinematical Hilbert space $\Hkin$ of CLQG.
The constraint \Ref{contA} means that the SU(2) part of the connection
$\SA$ is completely determined by the `momentum' variable $\tP$.
The latter is an operator and therefore, strictly speaking, it is not clear
in what sense it can define the rotational part. This problem is intimately
related to the issue of non-commutativity of the connection, which is one of the main
open problems of the present approach. In this work we consistently ignore this issue
and hence we accept the viewpoint that, due to the second class constraints, the projected
spin networks depend non-trivially only on the boost part of the connection.
Making a further simplification, we assume also that one can consider the holonomies
of such connection as spanning the homogeneous space $X\equiv {\rm SL}(2,\Cb)/{\rm SU}(2)$.
Then it is natural to expect that the necessary restriction on the labels is provided
by a reduction to simple representations. In other words,
it amounts to take all $n_k=0$ where the pairs $(n_k,\rho_k)$ label
irreducible representations of \SLC\ assigned to the edges.
We emphasize that this is only a conjecture that such a restriction is sufficient to remove all
degeneracy present in $\CH_0$.
We were not able to prove it in a rigorous way.

If this is a correct restriction, physical results should not depend on $n_k$.
Let us consider, for example, the spectrum of the area operator.
The contribution of an edge of a projected spin network parameterized by $(n,\rho)$ and $j_t,j_s$
near, say, the target vertex is given by \cite{AV}
\be
\CS \sim \sqrt{j_t(j_t+1)-n^2+\rho^2+1}.
\label{areasp}
\ee
Although it explicitly depends on $n$, it is clear that for any pair $(n,\rho)$ there exists
another pair $(0,\rho')$ which gives the same contribution to the area.\footnote{If $n>\rho$,
one should also adjust $j_t$ in an appropriate way.}
In this sense the dependence on $n$ is redundant.

Of course, a further investigation is needed to establish the consistency of the proposed restriction.
Here we suggest it as a working hypothesis. As a result, the kinematical Hilbert space $\Hkin$
of CLQG consists of projected spin networks with the following coloring (we use the second basis
of section \ref{subsec_enlarge}):
\begin{enumerate}
\item $(0,\rho_k)$ --- \SLC\ simple irreducible representations assigned to the edges;
\item $j_{t(k)}$ and $j_{s(k)}$ --- two SU(2) irreducible representations assigned to the edges;
\item $N_v$ --- SU(2) invariant intertwiners assigned to the vertices.
\end{enumerate}

In \cite{AlLiv} a further reduction was suggested. It is achieved by choosing
$j_{t(k)}=j_{s(k)}=0$. This choice is special since, after integration over $\chi$,
the resulting projected spin networks
reproduce simple spin networks appearing as boundary states of the BC spin foam model.
However, we do not see any reason form the canonical quantization point of view
to restrict ourselves to this particular set of states.
Moreover, one can observe that it is not closed under the action of various operators,
whereas the space $\Hkin$ is closed.
Therefore, it seems that the canonical quantization appeals for a generalization
of the BC model. And the first step in its construction is to generalize the unique
Barrett--Crane intertwiner which is at the heart of the model.

\section{Generalized Barrett--Crane intertwiners}

\subsection{Barrett--Crane intertwiner}

Before we present the intertwiners following from the canonical quantization
presented above, let us describe the unique intertwiner characterizing
the spin foam model of Barrett and Crane \cite{BCE,BC}. This will serve as a preparation
for future generalizations. Here we follow the
description of \cite{BFhigher}. The presentation can be done in a very general way
encompassing both Euclidean and Lorentzian cases in any dimension $d$.

Let us consider a group $G$ which is either the rotation group SO($d$)
or the Lorentz group SO($d-1$,1). We call an irreducible unitary
representation $\lambda$ of $G$ {\it simple} if its decomposition onto the
subgroup $H={\rm SO}(d-1)$ contains the trivial representation.
In the Lorentzian 4-dimensional case this definition amounts to consider representations
of the type $\lambda=(0,\rho)$.

The simple representations can be realized on functions on the homogeneous
space $X=G/H$. We denote $e^{(\lambda)}_p(x)$
an orthogonal basis in the representation space $H^{(\lambda)}_G$.
In particular, such a basis is given by the matrix elements of the group operators
\be
e^{(\lambda)}_p(x)=D^{(\lambda)}_{p0}(g_x),
\label{basise}
\ee
where the index $0$ corresponds to the basis element $e^{(\lambda)}_0$ invariant with respect to
the action of the subgroup $H$ (it gives rise to the trivial representation
appearing in the decomposition of the simple representation $\lambda$) and
$g_x$ is a representative of $x$ in the group $G$ (the matrix element does not depend
on its choice due to the invariance of $e^{(\lambda)}_0$).

An invariant intertwiner between a set of representations $\{\lambda_k\}_{k=1}^\LL$ is an operator
\be
N: \quad \bigotimes_{k=1}^\LL H^{(\lambda_k)}_G \to \Cb
\label{mapint}
\ee
which is invariant with respect to the action of $G$:
\be
N(\gl\cdot f_1,\dots,\gl\cdot f_\LL)=N(f_1,\dots,f_\LL), \qquad f_k\in H^{(\lambda_k)}_G.
\label{invNBC}
\ee
Its matrix elements are
\be
N_{p_1\dots p_L}=N(e^{(\lambda_1)}_{p_1},\dots,e^{(\lambda_\LL)}_{p_\LL}).
\ee

The Barrett--Crane intertwiner is characterized by an additional constraint.
Let $\hat T_{IJ}$ be generators of the Lie
algebra of $G$ ($I,J =1,\dots d$) acting on
functions on $X$, considered as a sphere or a hyperboloid in $\Rb^d$,
as derivatives
\be
\hat T_{IJ}\cdot f(x)=\eta_{IK} x^K \p_J f-\eta_{JK} x^K \p_I f.
\label{generator}
\ee
Here $x^I$ are coordinates in $\Rb^d$ and $\eta$ is either the unity matrix
or Minkowski metric depending on the group $G$.
Then the BC condition requires that for any pair of arguments the following equation
holds
\be
N_{\rm (BC)}( \hat T_{[IJ} \cdot f_1, \hat T_{KL]}\cdot f_2\dots,f_\LL)=0.
\label{constr}
\ee

In \cite{Reisenberger:1998bn} it was shown that there is a unique invariant
intertwiner, found first in \cite{BCE}, which satisfies \eqref{constr}.
It can be written in a very simple form
\be
N_{\rm (BC)}(f_1,\dots,f_\LL)=\int_X dx \, \prod_{k=1}^\LL f_k(x).
\ee
Taking into account \eqref{basise}, its matrix elements
are given by
\be
N_{{\rm (BC)}p_1\dots p_\LL}=\int_X dx \, \prod_{k=1}^\LL D^{(\lambda_k)}_{p_k 0}(g_x).
\label{matNBC}
\ee
The last equation is more convenient for generalizations which we are going to present
in the next subsection.

\subsection{Generalized intertwiner from the canonical approach}
\label{subsec_genint}

The analysis of section \ref{sec_impsec} implies that more general intertwiners should be considered
than the Barrett--Crane one. The latter appears only if one restricts all SU(2)
labels of projected spin networks
to $j=0$ and integrates over $\chi$ \cite{AlLiv,AK}. Otherwise one arrives at
the following generalization of the Barrett--Crane intertwiner
inspired by the Lorentz covariant approach to the loop quantization.

Let $N_H$ be an invariant intertwiner between $\LL$ representations of $H$
labeled by $j_k$. Accordingly to the decomposition
\be
H_G^{(\lambda)}=\bigoplus\limits_{j=0}^{\infty} H_H^{(j)},
\label{decompH}
\ee
let us split the range of values of the index $p$ into
sets corresponding to the representations of the subgroup,
$
p= (0,m_1, \dots ,m_j,\dots ),
$
so that $e^{(\lambda)}_{m_j}$ form a basis in $H_H^{(j)}$ appearing in the
decomposition \eqref{decompH}.
The decomposition starts with 0 because we consider $\lambda$ to be a simple representation.
Then we define the generalized intertwiner through its matrix elements
which are given by
\be
N^{(j_1,\dots, j_\LL)}_{p_1\dots p_\LL}(x)=\mathop{\sum}\limits_{m_{j_1},\dots,m_{j_\LL}}
N_H^{m_{j_1}\dots m_{j_\LL}} \prod_{k=1}^\LL D^{(\lambda_k)}_{p_k m_{j_k}}(g_x).
\label{genBC}
\ee

The proposed generalization is twofold. First, the projection to the trivial
representation presented in \eqref{matNBC}
has been replaced by a projection to a generic representation of
the maximal compact subgroup.
As a result, the new intertwiner \eqref{genBC}
is characterized not only by representations of $G$, which it couples,
but also by $\LL$ representations of $H$ appearing in the decompositions of
$\lambda_k$'s.
The independence of the representative $g_x$ is guaranteed
by the invariance of $N_H$, which replaced the trivial identity operator in
the expression for $N_{\rm (BC)}$.

Secondly, we dropped the integration over $x\in X$ so that the generalized intertwiner
is a function on the homogeneous space. As a result, it is not invariant anymore in the sense
of \eqref{invNBC}, but rather covariant satisfying the following property
\be
N(x;\gl\cdot f_1,\dots,\gl\cdot f_\LL)=N(\gl^{-1}\cdot x;f_1,\dots,f_\LL).
\label{invN}
\ee
Without this generalization, the first one does not have much sense. Indeed, the integration over $x$
makes an intertwiner invariant with respect to $G$. On the other hand, the intertwiners \eqref{genBC}
are only $H$-invariant and therefore the integration would produce simply an overcomplete basis
in the space of all $G$-invariant intertwiners.

The equation \eqref{invN} is the only condition satisfied by the generalized
intertwiners \eqref{genBC}. In other words, they form a basis in the space of
all maps \eqref{mapint} subject to the property \eqref{invN}.
In particular, the new intertwiners do not satisfy the property \eqref{constr} anymore,
which is a distinguishing feature of the Barrett--Crane intertwiner.
This property is a part of the so called simplicity constraints.
Classically, imposed on BF theory, they produce general relativity. The requirement
\eqref{constr} is an attempt to realize them at the quantum level,
where they are supposed to turn a topological spin foam model
of BF theory into a dynamical spin foam model of gravity.
Since the generalized intertwiners fail to satisfy this requirement,
it seems that we did not completely implement the simplicity constraints.
On the other hand, these constraints are nothing else but
the second class constraints \eqref{phi} taken into account
through the construction of Dirac brackets and the restriction to simple representations
on the Hilbert space. Therefore, there seems to be a contradiction between the two approaches.
To explain the apparent discrepancy we need to revise the
Barrett--Crane construction.

\section{Simplicity constraints and quantum 4-simplex}
\label{sec_spinfoam}

\subsection{Simplicity constraints in the spin foam approach}
\label{subsec_simplecon}

One of the ways to obtain the BC model is to start from a spin foam model of
the topological BF theory in 4 dimensions \cite{DePietri:1998mb,Freidel:1998pt}.
In this theory the independent fields are
a two-form $B$ valued in the adjoint representation of the Lorentz algebra
and a Lorentz connection $A$. The spin foam quantization in this case
is given by a Lorentzian version of the Crane--Yetter model \cite{Crane:1993if,Crane:1994ji}.
Its boundary states are given by spin networks with the Lorentz structure
group, {\it i.e.}, their edges are labeled by unitary irreducible Lorentz representations and
Lorentz invariant intertwiners are associated to their vertices \cite{FrLiv}.

At the classical level to pass from BF theory to general relativity,
one must ensure that the $B$ field is constructed from the tetrad one-forms, namely
\be
B=*(e\wedge e).
\label{Bee}
\ee
This is done by imposing the above mentioned simplicity constraints
\be
 B^{\alpha\beta}\wedge B^{\gamma\delta}= -{\cal V} \, \eps^{\alpha\beta\gamma\delta}
\ \Leftrightarrow\
\eps_{\alpha\beta\gamma\delta} B^{\alpha\beta}_{\mu\nu} B^{\gamma\delta}_{\rho\sigma}=
-{\cal V} \, \eps_{\mu\nu\rho\sigma} ,
\label{simplicityconditions1}
\ee
where $\CV= \frac{1}{4!}\,\eps_{\alpha\beta\gamma\delta} B^{\alpha\beta}\wedge B^{\gamma\delta}$
is the volume 4-form. To formulate these constraints at the quantum level, one uses
a natural identification of the $B$ field with the generators $\hat T$
of the gauge group, which are viewed as operators acting
on representation spaces associated with every face
of the spin foam (edge of the corresponding boundary spin network).
Then \eqref{simplicityconditions1} leads to various
restrictions on representations, intertwiners, {\it etc}.
In particular, when the two generators are associated to the same face, one gets
\be
\hat T\; \star\; \hat T=C_2=0.
\ee
Since the second Casimir operator for the Lorentz group is $C_2=2n\rho$, this leads to the restriction to
simple representations. On the other hand, when the two generators act on two different adjacent
faces, the corresponding quantum constraint is given by \eqref{constr}.
Its unique solution is the Barrett--Crane intertwiner.
Applied to the initial spin foam quantization of BF theory, these restrictions produce the BC model.

We believe that there is one weak point in this construction: the representation of the $B$ field by
the generators of the gauge group. It is definitely true in BF spin foam model and
the natural assumption is that it continues to hold in the constrained model so that it can be
used to formulate the constraints themselves. However, it seems to be inconsistent
with the canonical analysis.

To see this, let us notice that a way
to obtain the identification of $B$ with $\hat T$ is to consider an action of the operator
$\hat B$ on a holonomy of the connection $A$. In BF theory the $B$ field is canonically
conjugated to $A$ and therefore the action simply brings down the generator $T$ of the gauge group
\be
\int_S  \hat B\, \cdot\, U_{\gamma}[A]=-i\hbar\,U_{\gamma_1}[A]\cdot T\cdot U_{\gamma_2}[A],
\ee
where we assumed that a curve $\gamma$ is split into $\gamma_1 \cup\gamma_2$ by an intersection
with a surface $S$. More generally, as soon as
\be
[\, \eps^{jkl}\hat B_{kl}^{\alpha\beta}(y),A_i^{\gamma\delta}(x)]
=-i\hbar\,C^{\alpha\beta,\gamma\delta}\delta_i^j\delta(x,y)
\label{commAB}
\ee
with some arbitrary function $C^{\alpha\beta ,\gamma\delta}$, one obtains the same result with
$T^{\alpha\beta}$ replaced by $C^{\alpha\beta ,\gamma\delta} T_{\gamma\delta}$.
Thus, the identification is true only if $C^{\alpha\beta,\gamma\delta}=\eta^{\alpha[\gamma}\eta^{\delta]\beta}$,
{\it i.e.}, $B$ and $A$ are canonically conjugated.

On the other hand, the canonical analysis of the Plebanski formulation of general relativity shows
that it is not the case \cite{Buffenoir:2004vx,ABR}. The reason is that the stabilization procedure for
the simplicity constraints generates secondary constraints so that together
the simplicity and secondary constraints are second class.
As a result, the initial Poisson brackets are replaced by Dirac brackets which have a non-trivial form.
The analysis goes in the complete parallel with the covariant canonical analysis of
the Hilbert--Palatini action \cite{SA} where $\eps^{ijk}B_{jk}$ is identified with $\tP^i$ and
the role of the simplicity constraints is played by the constraints \eqref{phi}.
Similarly to that formulation, one can define a shifted connection
which satisfies \eqref{commAB} with the function $C^{\alpha\beta , \gamma\delta}$ given by the same
projector $I_{(P)}^{\alpha\beta, \gamma\delta}$ as in \eqref{comm}
\be
[\, \eps^{jkl}\hat B_{kl}^{\alpha\beta}(y),\SA_i^{\gamma\delta}(x)]
=-i\hbar\,I_{(P)}^{\alpha\beta, \gamma\delta}\delta_i^j\delta(x,y).
\label{commPleb}
\ee
Due to this, the identification
of $B$ with the generators holds only for the boost part of the field, whereas the rotation part
effectively vanishes.
This observation together with other critics of the BC model \cite{Livine:2007vk,Engle:2007uq}
suggests to look for an alternative to the Barrett--Crane quantization procedure.

\subsection{Quantizing a 4-simplex}
\label{subsec_new}

\subsubsection{Classical 4-simplex}

Deriving a new spin foam model, we will proceed in the way originally
used by Barrett and Crane \cite{BCE,BC}.
Namely, we are going to describe a `quantum 4-simplex' with a difference that now
we want to take into account lessons from the canonical quantization.

Let us consider an abstract 4-simplex in a 4-dimensional spacetime and
assign a bi-vector $b_k, \ k=1,\dots,10,$ to every triangle belonging to it.
This can be done by means of the integral\footnote{We inserted the Hodge operator into
the definition to make the bi-vectors coinciding with integrals of the $B$ field \eqref{Bee}.}
\be
b_k=\star\int_{\triangle_k} e\wedge e.
\label{defbiv}
\ee
As was proven in \cite{BCE}, at the classical level
the 4-simplex is completely characterized by $b_k$ provided they
satisfy a set of conditions. In particular, the most non-trivial conditions, which we are interested in, are
\begin{enumerate}
\item each bi-vector is simple, {\it i.e.} of the form $b=f\wedge g$;
\item if two triangles share a common
edge, then the sum of the corresponding two bi-vectors is also simple;
\item the sum of 4 bi-vectors corresponding to the faces of a tetrahedron is zero.
\end{enumerate}
The first two conditions are related to the simplicity constraints \eqref{simplicityconditions1}
on the $B$ field and the third one is called the closure constraint expressing the fact that
4 triangles must form a tetrahedron.

The next step is to realize these conditions at the quantum level where the bi-vectors are represented by
some operators. Whereas a solution of this problem was found by Barrett and Crane, here we argue
that there is another one, which is moreover consistent with the canonical quantization of general relativity.

\renewcommand{\labelenumi}{(\theenumi)'}

First, we notice that a classical 4-simplex can be also characterized in a different way.
Let us associate a normal vector $\vec x_v,\ v=1,\dots,5,$ to each of the 5 tetrahedra.
We will assume that all of them are timelike so that they can be viewed as elements
of the homogeneous space $X\equiv {\rm SL}(2,\Cb)/{\rm SU}(2)$. Any $x\in X$ gives rise to
two orthogonal projectors in the space of bi-vectors
\be
I_{(P)}^{\alpha\beta,\gamma\delta}(x)=-2x^{[\beta}\eta^{\alpha][\gamma} x^{\delta]},
\qquad
I_{(Q)}^{\alpha\beta,\gamma\delta}(x)=
\eta^{\alpha[\gamma}\eta^{\delta]\beta}+2x^{[\beta}\eta^{\alpha][\gamma} x^{\delta]}.
\label{projxxx}
\ee
The first one projects to those bi-vectors which are co-aligned with the vector $x^{\alpha}$
and the second projector selects bi-vectors orthogonal to it.
Then the 4-simplex can be uniquely determined by a set of $x_v$ associated to tetrahedra
and a set of bi-vectors $b_k$ such that
\begin{enumerate}
\item if the $k$th triangle belongs to the $v$th tetrahedron then $I_{(Q)}(x_v)b_k=0$;
\item the sum of 4 bi-vectors corresponding to the faces of a tetrahedron is zero:
$\smash{\sum\limits_{k(v)}}b_{k}=0$.
\end{enumerate}
Since according to our definition \eqref{defbiv} the bi-vectors appear as normals to triangles, the first condition
simply means that all triangles of a tetrahedron lie in a hypersurface orthogonal to $\vec x_v$.
In other words, associating an element of $X$ to each tetrahedron of the 4-simplex, we fix a spacelike
hypersurface which must contain all triangles of this tetrahedron.

A nice feature of this representation is that the simplicity constraints (the conditions (i) and (ii)
from the first list) are satisfied.\footnote{This is a direct consequence of the identity
$I_{(P)}^{\alpha\beta,\alpha'\beta'}(x)\eps_{\alpha'\beta'\gamma'\delta'}I_{(P)}^{\gamma'\delta',\gamma\delta}(x)=0$.
It ensures that the wedge product of any two bi-vectors satisfying
$b_k=I_{(P)}(x_v)b_k$, as follows from (i)', vanishes. The latter
property is well known to imply (i) and (ii).}
Thus, once a set of normals to tetrahedra is given, we provide
an explicit solution to these constraints. In this way one can avoid the problem of realizing the
simplicity constraints at the quantum level.

\renewcommand{\labelenumi}{(\theenumi)}

\subsubsection{Quantization}

Given the second characterization of a 4-simplex, we are ready to pass to its quantization.
First, we define a map from classical bi-vectors to quantum operators.
Using the isomorphism between the space
of bi-vectors and the Lie algebra $sl(2,\Cb)$, we represent a bi-vector $b_k$ on the representation space
assigned to the $v$th tetrahedron (which must, of course, contain the $k$th triangle) by the following
operator
\be
b_k\longmapsto \hat B_{k,v}= -i\hbar \,I_{(P)}(x_v)T_k,
\label{qmap}
\ee
where $T_k$ denote generators of $sl(2,\Cb)$. Some comments concerning this map are in order.

\begin{itemize}
\item
The operators carry two labels corresponding to a triangle and to a tetrahedron.
This is related to the fact that every such operator acts on representation spaces which
are attached to tetrahedra, rather than to triangles.
Since each triangle is shared by two tetrahedra, every bi-vector has two operator realizations.
In general, we only expect that these realizations can be related by a Lorentz transformation.
\item
The insertion of the projector ensures the fulfillment of the condition (i)' on bi-vectors.
It distinguishes our quantization from the usual one explored by Barrett and Crane
and can be thought as a particular way to solve the simplicity constraints. (Below we will discuss
more general solutions.)
\item
Identifying $x_v$ with the vectors \eqref{xxx} defined by the field $\chi$, one can check that
the projectors \eqref{projxxx} coincide with those of the canonical formulation \eqref{proj}.
Therefore, our quantization map \eqref{qmap} is in the precise agreement with the canonical
commutation relation \eqref{commPleb} valid in both Palatini and Plebanski formulations.
That commutator was found as a Dirac bracket taking into account the second class constraints.
Therefore, it is not surprising that a quantization consistent with this Dirac bracket
ensures the simplicity constraints.
\end{itemize}

Now we turn to the choice of representation spaces and intertwiners coupling them
at each tetrahedron. The set of representations to be considered must allow
to distinguish between inequivalent operators of type \eqref{qmap}.
Since such operators do not span the whole Lorentz algebra but only its boost part, it is
natural to expect that the simple representations are sufficient.
Thus, in the following we consider the representations of the type $(0,\rho)$ only.
Notice that we obtain the same restriction as in the BC model, but in a formally
different way. Although, of course, it is intimately related with the simplicity conditions.

The intertwiners associated to tetrahedra are usually fixed by the closure constraint,
which is the condition (ii)' from our list. Given the map \eqref{qmap}, it is written as
\be
\smash{\sum_{k(v)}}
\hat B_{k,v}N_v=0,
\label{clcon}
\ee
where $N_v\in \bigotimes\limits_{k(v)}H_G^{(\lambda_k)}$ is a vector in the tensor product
of simple representations representing an intertwiner.
This condition imposes the invariance of $N_v$ with respect to boosts. But since the commutator of
two boosts is a rotation operator, the full Lorentz invariance follows.
Thus, the condition \eqref{clcon} leads to boundary states which are the usual \SLC\
spin networks labeled by simple representations.

However, in our opinion, the condition \eqref{clcon} is a too strong requirement.
On the canonical quantization side, it would correspond to the integration of spin networks
over the field $\chi$. To obtain a spin foam model consistent with the loop quantization,
we relax the closure constraint to the following condition
\be
\smash{\sum\limits_{k(v)}}\hat B_{k,v}N_v(x_v)=i\hbar \,I_{(P)}(x_v)\hat T\cdot N_v\(x_v\),
\label{clconrel}
\ee
where $\hat T$ acts as in \eqref{generator}.
Thus, we allow the intertwiners to depend on $x_v$, which also transform under gauge transformations.
As a result, $N_v(x_v)$ are not invariant with respect to \SLC\ transformations, but only with respect to
the action of an SU(2) subgroup.
In section \ref{subsec_enlarge} we found two bases in the space of such intertwiners
and one of them was later explicitly presented in section \ref{subsec_genint} (eq. \eqref{genBC}).
Thus, we conclude that generalizing the closure constraint to the quantum case according to \eqref{clconrel},
one obtains the boundary states of a quantum 4-simplex which reproduce the projected spin networks of CLQG.

However, the generalization \eqref{clconrel} gives rise to a problem. What is the fate
of the classical closure constraint necessary for a quantum state to have a geometric interpretation?
At this stage we cannot answer this question, but rather refer to
two spin foam models, where a quite similar situation was found.

First, in \cite{Livine:2005tp} a model has been proposed,
which is obtained by generalizing the group field theory (GFT) \cite{Oriti,Freidel:2005qe}
for the BC model. The standard GFT is based on a field living on four copies of the group manifold
\cite{Perez:2000ec}, whereas the authors of \cite{Livine:2005tp} suggest to
consider a field with five arguments. The fifth argument is interpreted as the normal
to tetrahedra of the dual triangulation. The dependence on this additional
variable leads to intertwiners of precisely the type we deal with.
However, the actual model depends on the action one chooses for the field.
It was shown that there is one leading to the standard BC model, but also there are many others
which produce more general structures.\footnote{Let us notice that in this way can solve the `ultra-locality'
problem of the BC model, {\it i.e.}, one introduces a coupling between normals
to the same tetrahedron seen from different 4-simplices. A similar aim was pursued
in the recent attempts to modify the BC model \cite{Livine:2007vk,Engle:2007uq,Freidel:2007py}.}
In particular, it would be interesting to find
an action corresponding to our canonically motivated model. We leave this problem for
a future research.

The second model where one finds a similar deviation from the classical closure constraint
is the one proposed in \cite{Freidel:2007py}. That model is derived
by means of imposing the simplicity constraints on the discretized path integral for
BF theory rewritten in a special way. In contrast to other constraints, the closure constraint is not imposed
explicitly, but appears as a result of integration over group elements represneting holonomies
of a connection field. Therefore, although no discussion of boundary states appears in
\cite{Freidel:2007py}, one may expect that such states will spoil the condition \eqref{clcon}.
But, of course, a deeper analysis is needed to make a definite conclusion.

Finally, we remark that the r.h.s. of \eqref{clconrel} seems to be of quantum nature
since we do not know any classical quantity corresponding to it. Thus,
the quantum condition \eqref{clconrel} can be viewed as a microscopic generalization
of the classical relation $\sum b_k=0$, which is still valid at the macroscopic level.

\subsection{Alternative quantizations and relation to other models}
\label{subsec_alt}

\subsubsection{Other solutions and LQG}

\label{subsec_more}

In fact, the quantization proposed in the previous subsection is only one of
a two-parameter family of possible quantizations which solve the simplicity constraints.
The freedom comes from the possibility to `rotate' the isomorphism between bi-vectors
and the Lie algebra $sl(2,\Cb)$.
It is clear that for any values of $a$ and $b$, the following operators
\be
b_k\longmapsto \hat B_{k,v}(a,b)= -i\hbar \,I_{(P)}(x_v)\((1-b)+a\star\)T_k
\label{qmapgen}
\ee
satisfy the condition $I_{(Q)}(x_v)\hat B_{k,v}=0$ and thus may give a quantization
of a 4-simplex.\footnote{It is easy to realize that \eqref{qmapgen} is the unique solution
which transforms as a bi-vector under Lorentz transformations.
Indeed, the condition $I_{(Q)}\hat B=0$ implies that $\hat B=I_{(P)} {\Theta} T$
where $\Theta$ is an invariant tensor in $(\Lambda^2)^{\otimes 2}$. But there are only two invariant tensors
in this space: an antisymmetric combination of $\eta^{\alpha\beta}$ and the Levi--Civita symbol
$\eps^{\alpha\beta\gamma\delta}$ representing the action of the star operator.
Thus, one arrives at the linear combination \eqref{qmapgen}.}

The map \eqref{qmapgen} has a direct counterpart on the canonical quantization side.
Indeed, in \cite{SAcon} it was found that there is a family of Lorentz connections dependent of
two real parameters with the following commutation relation
\be
\{ \SA^X_i(a,b),\tP_Y^j\}_D=\delta_i^j \left(
(1-b)\delta^X_{X'} -a\Pi^X_{X'} \right) I_{(P)Y}^{X'}.
\label{abcom}
\ee
Applying our arguments from section \ref{subsec_simplecon} using this commutator,
one arrives precisely at the above given identification \eqref{qmapgen}.

However, in the same paper \cite{SAcon} it was argued that
the case $a=b=0$ is distinguished because it is the only one where
the corresponding connection has spacetime interpretation, {\it i.e.}, it transforms
correctly with respect to all gauge transformations and diffeomorphisms.
Other connections fail to transform appropriately under the action of the full Hamiltonian
generating time diffeomorphisms. This is the reason why we worked in this paper with
the connection \eqref{spconnew} which is precisely $\SA_i^X(0,0)$.
Nevertheless, here we want to examine the other cases as well.

For generic $a$ and $b$ it is not clear what restrictions one should impose on representations
associated to triangles to select a set which will be sufficient for the operators \eqref{qmapgen}.
The general case does not seem to produce a viable quantization.
Nevertheless, besides the above considered case $a=b=0$, there is another one which is special.
It is obtained when $b=1$ and the parameter $a$ is identified with the Immirzi parameter
according to $a=-\beta$.
For this particular choice of parameters,
the bi-vectors are represented by rotation generators
since \eqref{qmapgen} can be equivalently rewritten as
\be
b_k\longmapsto \hat B_{k,v}(-\beta,1)= i\hbar\,\beta \star I_{(Q)}(x_v)T_k.
\label{qmaprot}
\ee
Therefore, for a given tetrahedron $v$, the operators \eqref{qmaprot} form an SU(2) subgroup
and this fact is crucial for the quantization.

On the canonical side this case is also distinguished.
Actually, the connection $\SA^X_i(-\beta,1)$ is the same as $\SSA_i^X$ appearing in \eqref{su2con}.
In \cite{AlLiv} it was demonstrated that
this connection is an \SLC\ generalization of the Ashtekar--Barbero connection
and, in particular, it is commutative. Moreover, the loop quantization
based on the holonomies of this connection produces precisely the Hilbert space of LQG.
Thus, this gives a possibility to present LQG in a Lorentz covariant way where
the full \SLC\ symmetry is retained.
Here we show that the same result can be obtained using the map \eqref{qmaprot}.
This will provide a spin foam quantization consistent with the standard SU(2) LQG.

Once the quantization map for bi-vectors is given, it is enough to
determine an appropriate set of representations and intertwiners constrained
by (a generalization of) the closure condition.
Since the operators \eqref{qmaprot} belong to SU(2), we need
irreducible representations of only this group and the boundary states will be functions
on SU(2). On the other hand, to keep
the \SLC\ symmetry, one should embed the SU(2) representations into an \SLC\
irreducible representation. It is enough to choose one fixed \SLC\ representation
containing all SU(2) representations, which means that it must be simple.
Thus, to each triangle in the 4-simplex, one associates one representation of type $(0,\rho)$.

To get intertwiners to be considered, we impose an analogue of \eqref{clconrel}
where the r.h.s. is replaced by $-i\hbar\,\beta \star I_{(Q)}(x_v) \hat T \cdot N_v(x_v)$.
But it is easy to realize that this expression vanishes.
Therefore, the relaxed constraint coincides with the usual one \eqref{clcon}.
In our case the latter requires invariance of intertwiners only with respect to SU(2)
which is a stationary subgroup of $x_v$.
As a result, one arrives at the same space of intertwiners we discussed above
with a basis realized by the generalized intertwiners \eqref{genBC}.

This shows that the boundary states of a 4-simplex quantized according to \eqref{qmaprot}
are given by projected spin networks with fixed simple \SLC\ representations
on the edges. However, one should take into account that
the holonomies associated to the edges are constrained to lie in an SU(2) subgroup,
more precisely in $g_{x_{t(k)}}{\rm SU}(2)g_{x_{s(k)}}^{-1}$.
Then it is easy to realize that all projected spin networks with $j_{t(k)}\ne j_{s(k)}$ vanish
on this subspace and do not depend on the chosen \SLC\ representations $(0,\rho_k)$.
Thus, we remain with spin networks of the following form
\be
 \Psi_{j_k,i_v}\(\{h_k\}_{k=1}^{10},\{x_v\}_{v=1}^5\)=
\bigotimes\limits_{v=1}^5 N^{(\{j_{k(v)}\})}(x_v;i_v) \cdot
\bigotimes\limits_{k=1}^{10}
D_{\rm SL(2,\Cb)}^{(0,\rho_k)}\(g_{x_{t(k)}}h_k g^{-1}_{x_{s(k)}}\),
\label{su2spin}
\ee
where $h_k\in SU(2)$ and $N^{(\{j_{k(v)}\})}(x_v;i_v) $ are generalized intertwiners \eqref{genBC}
labeled by SU(2) spins $i_v$ every of which parameterizes the basis of SU(2)
invariant intertwiners between 4 representations.
However it is easy to see that the dependence on $x_v$ is canceled and one gets the usual
SU(2) spin networks. The representation \eqref{su2spin} is simply a way to
embedded them into a Lorentz covariant object. The embedding is controlled by $x_v$
which ensure the \SLC\ covariance.
This construction essentially repeats the one in \cite{AlLiv} and reproduces the Hilbert space of LQG.

Although we found a way to obtain a spin foam model consistent with LQG,
the physical interpretation of the quantization \eqref{qmaprot} is not transparent.
The presence of the Hodge operator in the quantization rule for bi-vectors
is in contradiction with the known identification for a 4-simplex \cite{Baez:1999tk}.
It implies that to quantize the theory, one uses not a canonical identification of bi-vectors with elements of
the gauge algebra, but a deformed one. In the langauge of \cite{Baez:1999tk} it corresponds to the use
of the non-flipped Poisson structure leading to `fake tetrahedra', whereas real tetrahedra follow from
the flipped one.
Moreover, the deformation \eqref{qmaprot} exists only in 4-dimensions
so that this procedure is not generalizable to other dimensions.
Thus, its physical explanation is not clear to us.
This may be considered as a counterpart of the above mentioned problem of the
corresponding connection $\SSA^X_i$
in the canonical approach where it fails to have the correct spacetime behavior.

\subsubsection{Relation to \cite{Engle:2007uq}}
\label{subsec_Rov}

There is a close relation of our work to the model suggested in \cite{Engle:2007uq}.
The authors considered a 4-simplex in Euclidean 4-dimensional space and proposed to quantize
it imposing the simplicity constraints in a weak sense. As a result, they obtained the usual restriction
to the simple representations on triangles of the 4-simplex and also, what is one of their main results,
an expression for the allowed intertwiners generalizing the Barrett--Crane one.
One can verify that they can be expressed in terms of our generalized intertwiners \eqref{genBC}
as integrals with respect to the normals to tetrahedra
\be
N^{(j_1,\dots, j_\LL)}_{{\rm (EPR)}p_1\dots p_\LL}=\int_X dx\,
N^{(j_1,\dots, j_\LL)}_{p_1\dots p_\LL}(x).
\label{interint}
\ee
Besides, the representations $j_k$ labeling the intertwiners were fixed
in terms of the simple SO(4) representations $(J_k,J_k)$
associated to triangles as
\be
j_k=2J_k.
\label{reprjj}
\ee
The resulting boundary states were claimed to reproduce
the SU(2) spin networks of LQG.

The correspondence \eqref{interint} tells us that the quantization of \cite{Engle:2007uq}
is closely related to our procedure.
In fact, looking how the intertwiners \eqref{interint} were derived, one immediately observes
that the following quantization map was used for bi-vectors
\be
b_k\longmapsto \hat B_{k}= i\hbar\,\star T_k.
\label{qmapr}
\ee
Thus, the Hodge operator was inserted comparing to the usual Barrett--Crane quantization.
A similar insertion was done also in \eqref{qmaprot}. Therefore, it is not surprising that
in both cases one arrives at a model which is very close to LQG.
The difference in the results of the two quantization procedures can be summarized as follows:
\begin{enumerate}
\item in the model \cite{Engle:2007uq} the SO(3) labels are determined in terms of
SO(4) representations \eqref{reprjj};
\item in our case the intertwiners are functions of $x_v$, whereas in \cite{Engle:2007uq}
they are integrated according to \eqref{interint}.
\end{enumerate}

In our opinion in both cases there are arguments in favor of our quantization
\eqref{qmaprot}.\footnote{Which is by itself has problems as discussed above.}
First of all, the first restriction cannot be translated to the Lorentzian theory. The \SLC\
representations are labeled by a continuous number $\rho$, whereas the label of SU(2)
representations is discrete.\footnote{There is also a discrete label in the representations of \SLC.
In particular, one could choose to consider another series of simple representations labeled
by $(n,0)$. But this series appears in the harmonic analysis on the homogeneous space
${\rm SL}(2,\Cb)/{\rm SL}(2,\Rb)$. It plays no role in LQG and therefore there are no reasons to
expect this series to be important as well.}
It is difficult to expect a relation between these two.
What however happens is that our states do not depend on the SO(4) representations $(J_k,J_k)$
and they can be fixed in arbitrary way satisfying $J_k\ge j_k$. In particular, one can choose
them as in \eqref{reprjj}. In this sense it is better to think that SO(4) labels are determined by
SO(3) representations.

On the other hand, the states and the vertex amplitude of \cite{Engle:2007uq}
seem to depend on the choice of $J_k$. This can be explained taking into account
the second distinction. Let us write the boundary states of \cite{Engle:2007uq}
considered as functions on SO(4) in the following form
\be
 \Psi_{j_k,i_v}\(\{g_k\}_{k=1}^{10}\)=
\bigotimes\limits_{v=1}^5 \int_X dx_v\,N^{(\{j_{k(v)}\})}(x_v;i_v) \cdot
\bigotimes\limits_{k=1}^{10}
D_{{\rm SO}(4)}^{(J_k,J_k)}\(g_k\).
\label{su2jj}
\ee
It is clear that it is not possible to get rid of the integrals over $x_v$ by whatever
choice of the group elements.
In our case the normals disappeared due to the special choice $g_k=g_{x_{t(k)}}h_kg^{-1}_{x_{s(k)}}$
(see \eqref{su2spin}).
But here it does not work due to the integration and therefore,
strictly speaking, the spin networks \eqref{su2jj}
cannot be reduced to the ordinary SU(2) spin networks.
This is also the reason why the dependence of $J_k$ survives.

Due to this, we conclude that these features indicate in favor of the spin foam quantization
which inserts projectors in the identification of bi-vectors with symmetry generators.
To our view, it allows to obtain results which are more consistent and coherent in both Euclidean
and Lorentzian signatures.

\subsection{Vertex amplitude}
\label{subsec_vertex}

Up to now we discussed only the kinematics. The dynamics of a spin foam model is encoded
in its vertex amplitude which is the quantum amplitude for a 4-simplex written in the spin network basis.
Thus, to complete the formulation of our model, we have to provide this crucial piece of information.

Usually, the vertex amplitude can be obtained by evaluating the boundary
spin network on the trivial connection. The result is then given by a coupling of
5 intertwiners assigned to the tetrahedra. However, as it is, this procedure does not work in our case.
Indeed, if one follows it, one obtains an amplitude which depends on $x_v$ and is not gauge invariant.

To understand the origin of the problem, one needs to return to the original definition of
the vertex amplitude and redo the calculations in our case. We borrow them from \cite{Engle:2007uq}.
Let us consider a single 4-simplex and discretize the action of BF theory on it. At this level
the $B$ field is represented by its values $B_k$ on triangles and the connection associates
a group element $\gl_v$ to every tetrahedron. Then the partition function for the single 4-simplex
with fixed $B_k$ on the boundary can be written as
\be
A(B_k)=\int \prod_v D\gl_v \, e^{i\sum_k \Tr\(B_k \gl_{t(k)}^{-1}\gl_{s(k)}\)}
=\int \prod_k g_k\, e^{i\Tr(B_k g_k)}\int\prod_v D\gl_v \,\delta\( \gl_{t(k)}g_k\gl_{s(k)}^{-1}\).
\ee
From this result we read the simplex amplitude in the connection representation whose scalar product with a boundary
spin network should produce the vertex amplitude. This gives
\be
A({\Upsilon})=\int \prod_k dg_k\, \Psi_{\Upsilon}(g_k)A(g_k)
=\int\prod_v d\gl_v\, \Psi_{\Upsilon}(\gl_{t(k)}^{-1}\gl_{s(k)}),
\label{vertusual}
\ee
where we denoted by $\Upsilon$ all collection of possible indices. The integration in \eqref{vertusual}
is in fact equivalent to imposing the closure constraint in its usual form \eqref{clcon}.
Therefore, if it was already imposed, the result is simply $\Psi_{\Upsilon}(\unity)$ as was announced above.
However, if only the weak version \eqref{clconrel} of the closure constraint was imposed,
the result is different. This explains why the usual prescription does not work in our case.
Then the amplitude is given by the spin network evaluated on the trivial connection and with
intertwiners integrated with respect to the normals $x_v$.
In particular, for the model discussed in section \ref{subsec_alt}, such integration
maps the spin networks to the form \eqref{su2jj} appearing in \cite{Engle:2007uq}
and hence the procedure described reproduces the vertex amplitude from that work.

Although this procedure allows to obtain nice non-trivial amplitudes, the result does not seem
to be satisfactory. It is inconsistent neither with the structure of the kinematical Hilbert space,
nor with considerations from canonical theory.
Let us concentrate for simplicity on the Euclidean version of the quantization from
section \ref{subsec_alt}. Similar considerations can be applied to our basic
Lorentzian model from section \ref{subsec_new}, but some reasonings become more involved in this case.

The first immediate problem arising for the quantization \eqref{qmaprot}
is that the amplitude following from the result \eqref{vertusual} depends on the spins $J_k$
labeling the SO(4) representations, whereas the elements of the kinematical Hilbert space
do not. In fact, the dependence in the spin networks disappears due to the restriction
of the holonomies to the subgroups $g_{x_{t(k)}}{\rm SU}(2)g_{x_{s(k)}}^{-1}$.
The necessity of this restriction is most explicitly seen in the canonical formalism
where the corresponding connection satisfies \eqref{su2con}.
This relation is the origin of all distinctions of the present approach from the standard
spin foam quantization.

In particular, it indicates that the way the vertex amplitude was derived should be modified
to take it into account. Indeed, in \eqref{vertusual} one integrates over the whole SO(4) group,
whereas the constraint \eqref{su2con} suggests that the integration must be restricted to
$g_{x_{t(k)}}{\rm SU}(2)g_{x_{s(k)}}^{-1}$.
If one does this, the vertex amplitude can be given by evaluation
of the boundary SU(2) spin network on the trivial connection producing the conventional
Wigner SU(2) 15J symbol.

Such a result is very suspicious because it corresponds to the topological SU(2) BF
theory.\footnote{We are grateful to the referee for pointing this out.}
Although in \cite{Freidel:2007py} it was argued that the model of \cite{Engle:2007uq},
which is very close to the model we discuss here, describes the {\it topological} sector of
Plebanski formulation,
we expect a more complicated vertex. In particular, since this model was constructed as
a spin foam cousin of LQG, it must encode matrix elements of the LQG Hamiltonian.
It is unlikely that they can be retrieved from such a simple vertex amplitude.

It seems that the derivation of the vertex amplitude we presented is too naive to take into account
the constraints \eqref{su2con} correctly. Our approach was the following:
first, we evaluated the simplex amplitude in BF theory and then restricted
it to the states of our model. However, as we saw on the example of the simplicity constraints,
this strategy can be dangerous. The constraints \eqref{su2con} are second class and must be taken
into account from the very beginning together with the constraints on the $B$-field.
Moreover, the correct path integral quantization requires a non-trivial measure
defined by the symplectic structure and commutation relations of the constraints.
As was discussed in \cite{Bojowald:2003im}, such measure factors are usually ignored
in the spin foam approach,
whereas they can essentially modify the amplitudes.

Thus, we have to leave the issue of the vertex amplitude open.
For its solution it will be crucial to provide a careful path integral
quantization of Plebanski formulation of gravity.
We hope to return to this important problem in future publications.

\section{Conclusions}

In this paper we suggested a new way to quantize a 4-simplex and to construct a spin foam model
for 4-dimensional general relativity. The construction is motivated by and is in agreement
with the canonical loop approach to quantum gravity or, more precisely, with its covariant version
known as CLQG. The main modification with respect to the Barrett--Crane procedure
concerns the assignment of operators to bi-vectors associated to triangles.
They are given by generators of the symmetry algebra with additional insertion of projectors
depending on the normals to tetrahedra. With such quantization rule, the simplicity constraints
are already satisfied and do not produce any further restrictions.

To get the complete agreement between the boundary states of the quantum 4-simplex and the kinematical
Hilbert space of CLQG, we had to relax the closure constraint at the quantum level.
Its new version allows intertwiners dependent of the normals. We noticed that there are already
a few spin foam models sharing this property, but a physical interpretation of this modification
is still to be revealed.

Besides, we found that there is a freedom in the proposed quantization of a 4-simplex, which is in a
nice agreement with the corresponding freedom on the canonical side.
In particular, there are two distinguished quantizations both in the canonical and spin foam frameworks.
The first leads to CLQG and the model we just described and the second produces a covariant
representation for the standard LQG with the SU(2) gauge group.

We also discussed the relation of our work to the model \cite{Engle:2007uq}. It turns out that
the states of that model can be obtained as particular combinations of projected spin networks,
which are the basic objects of our construction. At the same time, there are several discrepancies
related to a difference in the quantization procedure.

This work is only the first step to a consistent spin foam quantization of gravity.
Many details of the full construction are still lacking. For example, we provided only
the boundary states, whereas the discussion of the vertex amplitude was not conclusive.
We did not discuss at all the amplitudes for tetrahedra and triangles,
which can also contribute to the spin foam partition function.
It is likely that this will require a careful path integral derivation which takes
into account what we learnt in this paper.
Another very important issue is the large spin asymptotics of the spin foam amplitudes.
Besides its relevance to the semiclassical limit of the model, it is also essential for
understanding the geometric interpretation
of the quantum 4-simplex.

There are many unsolved problems on the canonical side as well. For example,
the way the second class constraints were imposed
on the Hilbert space is not rigorous and requires a better understanding.
In particular, although it seems very natural,
the restriction to simple representations in both canonical and spin foam quantizations
presented here is only a hypothesis up to now.
At the same time, the most non-trivial issue is given by the non-commutativity of the connection.
Its meaning for the loop quantization remains unclear, although it is likely that it will play an important
role in future investigations.

\section*{Acknowledgements}

The author is grateful to Jonathan Engle, Kirill Krasnov, Roberto Pereira, Alejandro Perez,
Philippe Roche, Carlo Rovelli and Hanno Sahlmann for very valuable discussions.
Also it is a pleasure to thank an anonymous referee whose comments allowed to improve
the presentation.
This research is supported by CNRS and by the contract ANR-06-BLAN-0050.

\end{document}